\documentclass{eptcs}
\providecommand{\event}{FMAS 2021} 
\usepackage{underscore}           
\usepackage{amsmath,amssymb}
\usepackage[capitalise, nameinlink, sort]{cleveref}
\usepackage{hyperref}

\usepackage{xspace}
\usepackage{multirow}
\newcommand{\CAN}{\textsc{Can}\xspace}
\newcommand{\ie}{\emph{i.e.}\@\xspace}
\newcommand{\eg}{\emph{e.g.}\@\xspace}
\makeatletter
\newcommand*{\etc}{\@ifnextchar{.}{etc}{etc.\@\xspace}}
\makeatother
\newcommand{\etal}{\emph{et al.}\@\xspace}
\usepackage{tikz}
\usetikzlibrary{automata,arrows,positioning,shapes}
\usepackage{todonotes}

\newcommand*{\defeq}{\stackrel{\text{def}}{=}}

\title{Observable and Attention-Directing BDI Agents for Human-Autonomy Teaming}
\author{Blair Archibald \qquad Muffy Calder \qquad Michele Sevegnani \qquad Mengwei Xu
\institute{School of Computing Science, University of Glasgow, Glasgow, UK}
\email{\{blair.archibald, muffy.calder, michele.sevegnani, mengwei.xu\}@glasgow.ac.uk}
}
\def\titlerunning{Observable and Attention-Directing BDI Agents for HAT}
\def\authorrunning{Blair Archibald et al.}
\begin{document}
\maketitle

\begin{abstract}

Human-autonomy teaming (HAT) scenarios feature humans and autonomous agents collaborating to meet a shared goal.
For effective collaboration, the agents must be transparent and able to share important information about their operation with   human teammates.
We address the challenge of transparency for Belief-Desire-Intention agents  defined in the  Conceptual Agent Notation (\CAN) language.  We  extend the semantics   to model agents that are \emph{observable} (\ie the internal state of tasks is available), and \emph{attention-directing} (\ie specific states can be flagged to users), and
  provide an executable semantics via an encoding in Milner's bigraphs.
Using an example of unmanned aerial vehicles, the BigraphER tool, and PRISM, we show and verify how the extensions work in practice.

\end{abstract}

\section{Introduction}\label{sec:intro}

Autonomous agents \eg robots~\cite{sauppe2015social} are becoming increasingly.
As the autonomy increases, so does the tasks that can be performed~\cite{seeber2020machines}, and more emphasis is being placed on how agents can form teams, alongside humans, to achieve a common goal, a.k.a. human-autonomy teaming (HAT)~\cite{shively2017human, o2020human}.

\emph{Transparency} is especially important in HAT.  While there is no single definition~\cite{winfield2021ieee, spagnolli2017transparency},  we draw on IEEE P7001 standard definition~\cite{P7001} that transparency is   measurable, testable, and is   “the transfer of information from an autonomous system or its designers to a stakeholder, which is honest, contains information relevant to the causes of some action, decision or behaviour and is presented at a level of abstraction and in a form meaningful to the stakeholder.”    Here, we focus on two transparency requirements:  \emph{observability} of internal states  and  \emph{attention-directing}   of a human teammate.

To illustrate transparency, consider a HAT scenario where a human  and an Unmanned Aerial Vehicle (UAV) survey    a site together. The UAV performs site surveys through imaging and retrieves objects if needed, while the human assembles and interprets   geographically related information (obtained from the UAV). 
To coordinate an effective exploration, it is essential to have the appropriate transparency into what tasks either is performing and the progress of the relevant tasks.
For example, the human needs to know if the UAV has finished surveying an area, or how much is left, before assembling analysis for the same area.
Transparency must also be supported to allow attention-direction during autonomy failure (\eg the engine is malfunctioning) and to highlight changes in the environment, \eg when a specific land condition  is detected by the UAV the human should be notified immediately for inspection.

We tackle the challenge of transparency by modelling and verifying agents that are \emph{observable}~\cite{bhaskara2020agent}, \ie the status of tasks is known, and \emph{attention-directing}, \ie specific states can be flagged to users.
Both observability and attention-directing are recommended features in a practical HAT system engineering guide~\cite{mcdermott2018human},
 though strictly this guide refers to the former as transparency and the latter as augmenting cognition, which we consider part of transparency.     
 
To demonstrate these features, we extend a model of Belief-Desire-Intention (BDI) agents proposed in~\cite{archibald2021modelling} that presents an executable semantics of the \CAN agent language~\cite{winikoff:declarative} based on Milner's Bigraphs~\cite{milner2009space}.
BDI agents~\cite{bratman1987intention, rao1995bdi} consist of, (B)eliefs: what the agent knows; (D)esires: what the agent wants to bring about; and (I)ntentions: the desires the agent is currently acting upon.
BDI agents are chosen as they are represented in a declarative fashion (easy to log) with some self-explanatory concepts \eg the beliefs of the agent.
\CAN is chosen as it features a high-level agent programming language that captures the essence of BDI concepts without describing implementation details such as data structures.
As a superset of AgentSpeak~\cite{rao1996agentspeak}, \CAN includes advanced BDI agent behaviours such as reasoning with \emph{declarative goals}, \emph{concurrency}, and \emph{failure recovery}, which are necessary for our UAV example modelled in~\Cref{sec:casestudy}.
Importantly, although we focus on the  \CAN, the language features are similar to those of other mainstream BDI languages and the same modelling techniques would apply to other BDI programming languages.
Besides the work of~\cite{bremner2019proactive}  that focuses on transparent ethical reasoning (\ie why a decision has been made by an agent), we believe this is the first formal analysis of transparency applied to mainstream BDI agents.

We make the following research contributions: (1) an extension of \CAN language semantics to support transparency, (2) an extension of the bigraph based executable semantics framework, (3) an evaluation based on Unmanned Aerial Vehicles (UAVs) to illustrate the framework.

\section{Framework}\label{sec:maincontent}

In this section, we provide transparency mechanisms for \emph{observability}, \ie documenting the status of tasks and their progress toward completion, and \emph{attention-directing}, \ie flagging information about specific states---of the situation or agent itself---to a user. We   start with some \CAN language background.

\subsection{\CAN Background}
	The \CAN language formalises a classical BDI agent  consisting of a belief base~$ \mathcal{B} $ and a plan library~$ \Pi $. The belief base $ \mathcal{B} $ is a set of formulas encoding the current beliefs and has belief operators for entailment (\ie $ \mathcal{B} \models \varphi ) $, and belief atom addition (resp. deletion)
	$ \mathcal{B} \cup \{b\}$ (resp. $\mathcal{B} \setminus \{b\} $) and any logic over the belief base $ \mathcal{B} $ is allowed providing entailment is supported.
	A plan library $ \Pi $ is a collection of plans of the form $ e : \varphi \leftarrow \mathit{P} $ with $ e $ the triggering event, $ \varphi $ the context condition, and $ \mathit{P} $ the plan-body. Events can be either be external (\ie from the environment in which the agent is operating) or internal (\ie sub-events that the agent itself tries to accomplish).  In the plan-body, we use $ \mathit{P}_{1};\mathit{P}_{2} $ for sequence and $ \mathit{goal}(\varphi_{s}, \mathit{P}, \varphi_{f}) $ for the declarative goal failing if $ \varphi_{f} $ holds and exiting successfully if $\varphi_{s}$ holds (see~\cite{sardina:goals}).

	A basic configuration $ \langle\mathcal{B}, P\rangle $, where $P$ is the plan-body being executed (\ie the current intention),  is used in rules that define the execution of a single intention.
    The agent configuration is defined as $\langle E^{e}, \mathcal{B}, \Gamma\rangle$ where $E^{e}$ denotes the a set of pending external events and $\Gamma$ the current set of intentions (\ie partially executed plan-body programs).
	The agent-level evolution is specified by the transitions over the agent configuration.
	For example, the \emph{agent-level} transition to progressing intention which is progessable ($\langle \mathcal{B},  P\rangle \rightarrow \langle  \mathcal{B}',  P'\rangle$) or dropping \emph{any} unprogressable intention ($\langle \mathcal{B},  P\rangle \nrightarrow$) can be given as follows:
	\begin{center}

		$ \dfrac{P \in \Gamma \ \ \ \langle \mathcal{B},  P\rangle \rightarrow \langle  \mathcal{B}',  P'\rangle }{\langle E^{e},  \mathcal{B},  \Gamma\rangle  \Rightarrow \langle E^{e}, \mathcal{B}', (\Gamma \setminus \{P\})\cup \{P'\}    \rangle } \  A_{step} $\qquad  
		$ \dfrac{P \in \Gamma \ \ \ \langle \mathcal{B},  P\rangle \nrightarrow}{\langle E^{e}, \mathcal{B},  \Gamma \rangle  \Rightarrow \langle E^{e}, \mathcal{B},  \Gamma \setminus\{P\}    \rangle }  A_{update} $
	\end{center}
	We refer the reader to \cite{winikoff:declarative, sardina:goals} for a full overview of the semantics of \textsc{Can}.

\subsection{Observability}\label{sec:observability}

Observability captures \emph{what} an agent is doing so that human teammates can coordinate their tasks for effective collaboration. As intention represents  tasks, we focus on the \emph{status} and \emph{progress} of intentions.

\subsubsection{Intention Status}

\begin{figure}

\begin{center}
    		$ \dfrac{e \in E^{e}}{\langle E^{e},  \mathcal{B},  \Gamma\rangle  \Rightarrow \langle E^{e} \setminus\{e\}, \mathcal{B},  \Gamma \cup \{e\}    \rangle } A_{event}   $
    		
    		$ \dfrac{\langle e, I, \texttt{pending}\rangle \in E^{e}}{\langle E^{e},  \mathcal{B},  \Gamma\rangle  \Rightarrow \langle E^{e} \setminus\{\langle e, I, \texttt{pending}\rangle\}\cup \{\langle e, I, \texttt{active}\rangle\}, \mathcal{B},  \Gamma \cup \{\langle e, I \rangle\}    \rangle } A^{new}_{event} $

\end{center}

\begin{center}
    		$ \dfrac{P \in \Gamma \ \ \ \langle \mathcal{B},  P\rangle \rightarrow \langle  \mathcal{B}',  P'\rangle }{\langle E^{e},  \mathcal{B},  \Gamma\rangle  \Rightarrow \langle E^{e}, \mathcal{B}', (\Gamma \setminus \{P\})\cup \{P'\}    \rangle }   A_{step}  $
    		
    		$  \dfrac{\langle P, I\rangle \in \Gamma \ \ \ \langle \mathcal{B},  \langle P, I\rangle \rangle \rightarrow \langle  \mathcal{B}',  \langle P', I\rangle \rangle }{\langle E^{e},  \mathcal{B},  \Gamma\rangle  \Rightarrow \langle E^{e}, \mathcal{B}', (\Gamma \setminus \{\langle P, I\rangle\})\cup \{\langle P', I\rangle\}    \rangle }   A^{new}_{step} $
    		
\end{center}

\begin{center}
   $\dfrac{P \in \Gamma \ \ \ \langle \mathcal{B},  P\rangle \nrightarrow}{\langle E^{e}, \mathcal{B},  \Gamma \rangle  \Rightarrow \langle E^{e}, \mathcal{B},  \Gamma \setminus\{P\}    \rangle }  A_{update}$
   
   $ \dfrac{\langle P, I\rangle \in \Gamma \ \ \ \langle e, I, \texttt{active}\rangle \in E^{e} \ \ \ \langle \mathcal{B},  \langle P, I\rangle \rangle \nrightarrow  \ \ \ P = nil    }{\langle E^{e}, \mathcal{B},  \Gamma \rangle  \Rightarrow \langle E^{e}\setminus\{\langle e, I, \texttt{active}\rangle\}\cup \{\langle e, I, \texttt{success}\rangle\}, \mathcal{B},  \Gamma \setminus\{\langle P, I\rangle\}    \rangle }  A^{new}_{update\_suc}   $
   
   $\dfrac{\langle P, I\rangle \in \Gamma \ \ \ \langle e, I, \texttt{active}\rangle \in E^{e} \ \ \ \langle \mathcal{B},  \langle P, I\rangle \rangle \nrightarrow  \ \ \ P \neq nil    }{\langle E^{e}, \mathcal{B},  \Gamma \rangle  \Rightarrow \langle E^{e}\setminus\{\langle e, I, \texttt{active}\rangle\}\cup \{\langle e, I, \texttt{failure}\rangle\}, \mathcal{B},  \Gamma \setminus\{\langle P, I\rangle\}    \rangle }  A^{new}_{update\_fail}  $

\end{center}

	\caption{Derivation rules for agent configuration.}
	\label{tbl:agentCANSemantics}
\end{figure}

As a high-level planning language, \CAN is agnostic to many practical issues --  including transparency.
One issue of \CAN is the inability to tell the status of an intention.
For example, the rule $A_{update} $ in \cref{tbl:agentCANSemantics} discards an intention that cannot do any more intention-level transitions: both if it has already succeeded, or if it failed, and there is no way to determine which was the case. In practice,   we may find it helpful, particularly in HAT, to have   precise knowledge on the success or failure status of an intention.

We extend \CAN semantics to allow intention status.
Following work~\cite{harland2014operational}, we introduce four status values for an external event (that ultimately gives rise to an intention): \texttt{pending}, \texttt{active}, \texttt{success}, and \texttt{failure} along with an unique identifier $I$. These values  indicate when an event is not addressed yet (\texttt{pending}), currently being addressed (\texttt{active}), successfully addressed (\texttt{success}), and  addressed with a failure (\texttt{failure}). Unique identifiers  enable the agent to track the means-end relations between the events and the related intentions, and differentiate intentions from others.

\Cref{tbl:agentCANSemantics} presents the original rules and our status-enabled rules where $ \langle E^e, \mathcal{B}, \Gamma\rangle $
consists of a set of external events $ E^{e} $ that the agent is required to respond, a belief set $\mathcal{B}$, and intention base $ \Gamma $ is a set of partially executed plan-bodies $ P $
that the agent has committed to.
While the original rule $ A_{event} $ deletes events once they have turned into intentions, the new rule $ A^{new}_{event} $ instead switches the status of the event to \texttt{active}.
When intentions progress, the status of related external events remains \texttt{active}, until   the intention is removed either with \texttt{sucess} (if we reached a $nil$) via $A^{new}_{update\_suc}$ or \texttt{failure} via $A^{new}_{update\_fail}$.

\subsubsection{Intention Progress}

Besides the intention status, it is also useful to estimate the progress of intentions.
For example, the human teammate may want to know how long the UAV needs to finish  surveying to plan their next tasks.

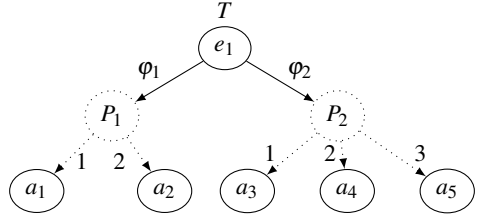
\begin{figure}
	\centering
	\footnotesize

	\begin{tikzpicture}[node distance=4.2em]
	\tikzstyle{ornode}=[draw, ellipse]
	\tikzstyle{andnode}=[draw, ellipse, dotted]
	\tikzstyle{oredge}=[->, -latex]
	\tikzstyle{andedge}=[->, dotted, -latex]
	\tikzstyle{andangle}=[draw, dotted, angle radius=1.3em]

	\node[ornode] (G3) at (-1.6, 0) {$ e_{1}$};
	\node[andnode] (P1) at (-3.1, -0.9) {$P_{1}$};
	\node[ornode] (a1) at (-4.1, -1.9) {$a_{1}$};
	\node[ornode] (a2) at (-2.4, -1.9) {$a_{2}$};

	\node[andnode] (P2) at (-0.1, -0.9) {$P_{2}$};
	\node[ornode] (b1) at (-1.3, -1.9) {$a_{3}$};
	\node[ornode,  right of=b1] (b2) {$a_{4}$};
	\node[ornode,  right of=b2] (b3) {$a_{5}$};

	\draw[oredge] (G3) edge (P1);
	\draw[andedge] (P1) edge (a1);
	\draw[andedge] (P1) edge (a2);

	
	\draw[oredge] (G3) edge (P2);
	\draw[andedge] (P2) edge (b1);
	\draw[andedge] (P2) edge (b2);
	\draw[andedge] (P2) edge (b3);

	\node[](1) at (-2.6, -2) {};
	\node[](2) at (-2, -2) {};
	
	\node[](3) at (-2.6, -2.5) {};
	\node[](4) at (-2, -2.5) {};

		
	
	\node[] at (-1.6, 0.5) {$ T$};

	\node[] at (-2.6, -0.3) {$\varphi_{1}$};
	\node[] at (-0.6, -0.3) {$\varphi_{2}$};

	\node[] at (-3.5, -1.5) {$ 1$};
	\node[] at (-3, -1.5) {$ 2$};

	\node[] at (-1, -1.4) {$ 1$};
	\node[] at (-0.2, -1.4) {$ 2$};
	\node[] at (1, -1.4) {$ 3$};

	\end{tikzpicture}
	\caption{A goal-plan tree.}
	\label{fig:graphforexplainingtraces}
\end{figure}

In BDI agents, goal-plan trees (shown in \cref{fig:graphforexplainingtraces}) are a canonical representation of intentions~\cite{thangarajah:computationally}. 
The root of the tree is an external event, and its children are plans that can handle this event.
Plans may contain sub-events, giving rise to a tree structure that represents all possible ways of achieving a task. 
Traditionally, for simplicity, goal-plan trees often do not represent actions and contain only goals/subgoals and plans \eg in~\cite{thangarajah:computationally}. However, we need to include the actions---a crucial part in agent execution---to perform  faithful intention progress estimation.
Finally, executing a BDI program gives \emph{one} execution trace\footnote{For practicality, we do not allow recursive plans as this can lead to infinite traces.}~\cite{xu2019intention}, \ie a path through the tree. 

To illustrate goal-plan trees and execution traces, consider the following two plans as follows:
\begin{center}
    $ P_{1} = e_{1} : \varphi_{1}   \leftarrow a_{1}; a_{2}; $ \ \ $ P_{2} = e_{1} :\varphi_{2}   \leftarrow a_{3}; a_{4}; a_{5}. $
\end{center}
These two plans are visualised as the goal-plan tree $ T $ given in~\cref{fig:graphforexplainingtraces}. The goal-plan tree $T$ expresses that, given an event $e_1$, it has two plan choices, namely plan $P_1$ and $P_2$, where the edges between the event $e_1$ and plan $P_1$ and $P_2$ are labelled with the relevant context condition $\varphi_{1}$ and $\varphi_{2}$. 
If plan $P_1$ is selected (given $\varphi_{1}$ holds), it has two sequenced actions $a_1$ and $a_2$ to execute where the edge between the plan to its plan-body is labelled with the position number.  As such,  we can obtain one  execution trace   $e_{1}; P_{1}; a_{1}; a_{2}$  and  the other one  (due to the selection of plan $P_2$) $e_{1}; P_{2}; a_{3}; a_{4};  a_{5}$.

For intention progress estimation, we use a compile-time process to obtain all possible execution traces.
For each trace, we record the maximum length of the execution trace.
During execution, the agent tracks its current execution trace so that after an agent-step we can determine the maximum length of the full trace this current trace belongs to\footnote{To ensure each trace is uniquely identifiable, the same action occurring in a different plan is treated as a different action.}. 
The progress of an intention is then estimated as the ratio of the position of the last element in the current trace and the length of the full trace this current trace belongs to. For example, given a trace $e_{1}; P_{1}; a_{1}; a_{2}$, its maximum length is 4 and the position for each element in this trace is  1 for $e_1$, 2 for $P_{1}$, 3 for $a_{1}$, and 4 for $a_{2}$. 
If the current execution trace is $e_{1}; P_{1}; a_{1}$ then the progress estimation is $3/4 = 75\%$ (as $a_1$ is the last element). In the case of failure recovery, for example, if action $a_{1}$ failed, the agent backtracked to event $e_1$ and subsequently  selected the plan $P_2$. Then the current trace becomes $e_{1}; P_{2}$ (against the full trace $e_{1}; P_{2}; a_{3}; a_{4};  a_{5}$) with the progress estimation $2/5 = 40\%$.  The execution trace is maintained separately for each intention.

\subsection{Directing Attention}\label{sec:directingattention}

Agents should be able to direct the attention of teammates when particular states become relevant, \eg when adversarial situations are encountered.
In BDI programming languages such as AgentSpeak, a change in the belief base   generates an event, which subsequently requires an agent to select a plan. For example, if a belief atom $b$ is true, an event in the form of $ +b$ is added to the  desires.  While useful,  this one-to-one mapping between belief and event is also limiting, especially in the context of directing attention \eg it prohibits the generation of multiple events. Further, multiple events generated by a single belief change may require adjustment throughout the operational life of the agent.

We take an approach to directing attention that  employs the motivation library of~\cite{sardina2011bdi}. This  allows new (and multiple) intentions (\eg to communicate with a human)  to be created dynamically when a particular state is recognised (\ie the correct beliefs hold).
To be precise, a motivation library (specified by the agent designers) $ \mathcal{M} $ is a set of rules of the form: $ \psi \rightsquigarrow \langle e, I\rangle $ where  $\psi$ is a world state, $e$ an event, and $I$ an identifier.
The semantics of motivation execution is specified as follows:
\begin{center}
 $ \dfrac{\psi \rightsquigarrow \langle e, I\rangle  \in \mathcal{M} \ \  \mathcal{B} \models \psi \ \ \langle e, I\rangle  \notin \Gamma \ \ }{\langle E^{e}, \mathcal{B},  \Gamma\rangle  \rightarrow \langle E^{e} \cup \langle e, I, \texttt{active}\rangle , \mathcal{B},  \Gamma\cup \{\langle e, I\rangle\}\rangle }$ \ \ $ A_{motive}$
\end{center}
Informally, if the agent believes $\psi$, it should adopt the event $\langle e, I\rangle$ (if it has not adopted it before).
As such, the programmers can specify (and revise) multiple rules $\psi \rightsquigarrow \langle e_1, I_1\rangle$, $\ldots$, $\psi \rightsquigarrow \langle e_n, I_n\rangle$ to ensure a correct set of events to be generated at any time when $\psi$ holds. From our experience, this approach also benefits from the modularity principle by separating the dynamics of desires and design of plan library.

\section{UAV Example}\label{sec:casestudy}


  \begin{figure}
  \small
		\begin{align*}
		&\texttt{1 \ // Initial beliefs}\\
		 &\texttt{2 \  $ \neg  $sensor\_malfunc, $ \neg  $engine\_malfunc}\\
		 &\texttt{3 \  // External events}\\
		 &\texttt{4 \  $ \langle$e\_retrv, identifier1$\rangle$}\\
		 &\texttt{5 \  // Motivation library events}\\
		 &\texttt{6 \  parked <- $ \langle$e\_parked, identifier2$\rangle$}\\
		 &\texttt{7 \  // Plan library}\\
		 &\texttt{8 \ \ e\_retrv : $ \varphi $ <- take\_off; goal(at\_destination, e\_path1, fc); retrieve}\\
		 &\texttt{9 \ \ e\_retrv : $ \varphi $ <- take\_off; goal(at\_destination, e\_path2, fc); retrieve}\\
		&\texttt{10 \ e\_retrv : $ \varphi $ <- take\_off; goal(at\_destination, e\_path3, fc); retrieve}\\
		 &\texttt{11 \ e\_retrv : sensor\_malfunc <- return\_base}\\
		 &\texttt{12 \ e\_retrv : engine\_malfunc <-  activate\_parking}\\
		 &\texttt{13 \ e\_path1 : true <-  navigate\_path\_1}\\
		&\texttt{14 \ e\_path2 : true <-  navigate\_path\_2}\\
		 &\texttt{15 \ e\_path3 : true <-  navigate\_path\_3}\\
		 &\texttt{16 e\_parked : true <-  send\_GPS}\\
		 &\text{where} \ \varphi  = \neg\texttt{sensor\_malfunc}  \wedge  \neg\texttt{engine\_malfunc},\\
		 & \texttt{fc} =  \texttt{sensor\_malfunc} \vee \texttt{engine\_malfunc} 
			\end{align*}
	\caption{UAV BDI agent design.}
	\label{fig:retrievetask}
\end{figure}

To demonstrate our new transparency mechanisms, we give a simple Unmanned Aerial Vehicles (UAVs) example, where  
UAVs are used for object retrieval tasks, \eg package delivery, and might be subject to engine or sensor malfunction.   The  agent design     is in~\cref{fig:retrievetask} where we heavily rely on declarative goal and failure recovery features in \CAN.
 There is one main retrieve task, initiated by external event \texttt{e\_retrv} (line 4), which can be handled by five relevant plans (lines 8--12).
The first 3 plans provide different flight paths after take-off in which they all have a declarative goal. For example, the declarative goal \texttt{goal(at\_destination, e\_path1, fc)} says that it is achieved if it believes \texttt{at\_destination} holds and failed if \texttt{fc} holds. When \texttt{fc} holds, the other two plans (lines 11 and 12) perform safe recovery in the event of engine or sensor malfunction. 
In the case of engine malfunction, instead of returning to base (when sensor is faulty), the plan in line 12   instructs the UAV to   land and park itself. Once the UAV is landed and parked, the human teammate should be notified of the situation and respond accordingly. Therefore, for directing attention, the motivation library is given in line 6, so that if the belief   that the UAV parked itself  (\ie \texttt{parked})  holds after the action \texttt{activate_parking}, the agent should adopt the event \texttt{e\_parked} to send GPS coordinates (line 16) to a human teammate to retrieve the UAV. For succinct presentation, we do not show the encoding of actions such as \texttt{take\_off} and \texttt{retrieve} (which can found in the  online 
model\footnote{\url{https://bitbucket.org/uog-bigraph/observable_attention-directing_bdi_model/src/master/}}).

\subsection{Analysis}

We    encoded the \CAN semantics in bigraphs~\cite{milner2009space,archibald2021modelling}, which
has  been  used previously to encode the  semantics of process calculi~\cite{bundgaard2006typed, memo2014}. Our  bigraph encoding permits execution  and symbolic analysis and we 
  employ BigraphER~\cite{bigraphER}---an open-source language and toolkit for bigraphs---to generate (and export) a transition system for analysis with model checking tools \eg PRISM~\cite{KNP11}. The example above  is  available in BigraphER format in the online model.
For reasoning, 
states are labelled with  \emph{bigraph patterns}~\cite{benford2016lions},      
predicates that indicate if  there is  match of a bigraph  in that state. 
Temporal properties are expressed using linear or branching time temporal logics \eg Computation Tree logic (CTL)~\cite{clarke1981design}.

For observability, we want to check      
1) if an intention is being progressed, its status should never be \texttt{pending},
2) if an intention becomes a completed empty program,   its related event  will eventually succeed, and
3) if an intention becomes blocked, but is not an empty program, its related original event   will eventually  fail.   
For progress,  we label each step in the execution trace  with its true progress estimation and check that there always exists a path where these match. 
For  attention-directing, we check that if the belief atom \texttt{parked} is believed, then eventually the event \texttt{e\_parked} will be added to the agent desires.

As an example, consider   observability and properties 2) and 3) for the event, namely \texttt{$ \langle$e\_retrv, identifier1$\rangle$} given in~\cref{fig:retrievetask}, which we can express in CTL as   follows:
\begin{center}
$ \mathbf{A} [   \varphi_{1}  \implies \mathbf{F} (\varphi_{2} \wedge \neg \varphi_{4})]\qquad\qquad$ for property 2) \\
$\mathbf{A} [ \varphi_{3}  \implies \mathbf{F} (\varphi_{4} \wedge \neg \varphi_{2}) ]\qquad\qquad$   for property 3)
\end{center}
The specification for property 2) checks that along all paths, if an intention is completed (\ie~$\varphi_{1}$ holds), this implies that  
eventually the original event will succeed (and not fail), \ie~$\mathbf{F} (\varphi_{2} \wedge \neg \varphi_{4})$. Similar logic applies when an intention is blocked for property 3).  We do not give full details, but for the interested reader, the state formulae can be represented by the following bigraph patterns:
\begin{center}
   $ \varphi_{1} \defeq \mathsf{Intent_{e}.(Identifier.Identifier1 \mid Nil \mid id)}$\\
   $ \varphi_{2} \defeq \mathsf{Event_{e}.(Identifier.Identifier1 \mid Success \mid id)} $, \\
 $ \varphi_{3} \defeq \mathsf{Intent_{e}.(Identifier.Identifier1 \mid ReduceF \mid id)}$ \\
 $ \varphi_{4} \defeq \mathsf{Event_{e}.(Identifier.Identifier1 \mid Failure \mid id)}$,
\end{center} where $\mathsf{Identifier1}$ is the identifier of the event \texttt{$ \langle$e\_retrv, identifier1$\rangle$}, the symbol $\mathsf{id}$ (called a site in bigraphs) stands for the part of model that is abstracted away such as the execution trace, and the subscript $\mathsf{e}$ (called a link)   maintains the mean-end relation between an original event and the related intention. 
The transition system for the agent in~\cref{fig:retrievetask} has  \textbf{44} states and \textbf{44} transitions\footnote{There are numerous  internal states, that do not appear in the final transition system, but add to build time.} and all the above mentioned properties are shown to hold.


\section{Related Work}\label{sec:related}

We are not the first to label intention status~\cite{harland2014operational,harland2017aborting} (as pending/active), however like the original \CAN semantics, these approaches do not make intention success/failure explicit. We believe having this information is much more important to human users than simply knowing an intention is dropped. Intention estimation has also been explored~\cite{thangarajah2014quantifying,thangarajah2015estimating}, but in the context of scheduling where they wish to choose the most-completed intention first. The approach is based on \emph{completeness} measures that use the resource consumption and effects of achieving intentions as an estimate, however this requires domain knowledge on the pre-conditions/post-effects of events that can be hard to obtain. For the attention directing feature, we adopted the motivation library approach from \cite{sardina2011bdi}. There are similar approaches such as conditional goals   in~\cite{van2004dynamics, van2005semantics} and automatic events~\cite{winikoff:jack} (adopting a goal or an event that is conditionalised by beliefs).

 Verifying BDI agents through model checking has been well explored, \eg \cite{dennis2016practical} verifies the decision making part (modelled in BDI agents) of a hybrid autonomous system,
 and~\cite{dennis2012model} uses model checking to reason about  agent programs written    in
   Agent Programming Languages. 
Bremner \etal verify
that BDI-based decision making (\eg plan selection) in robotic systems adhere to ethical rules (\eg save human)~\cite{bremner2019proactive}. For transparency, they propose a recorder that produces human readable logs consisting of information such as belief base and the plans executed. Though we agree on the use of logging as a means of transparency, their aim is to provide logs for future forensic analysis whereas ours is to provide run-time logs for effective HAT.

\section{Conclusion and Future Work}\label{sec:conclusion}

We presented an executable framework for verifiable BDI agents supporting \emph{observability} (showing task status) and \emph{attention-directing} (flagging relevant information), to facilitate effective human-autonomy teaming (HAT).
Observability allows comprehending \emph{what} an agent is doing and \emph{how much} progress has been made, while attention directing ensures communication of critical information to human teammates.

Observability is implemented by enabling the agent to show the status of intentions (\eg pending or active), and to provide a quantitative estimation of progress at each step.
Attention-directing is achieved through a motivation library that allows agents to adopt multiple new events when a particular world state is believed to hold.
Using a UAV example, we have demonstrated these features in practice.

Full scale observability and attention directing is beyond this initial framework.
Future work includes an in-depth survey to determine what are the most important aspects   to be made observable (to humans).
This might require domain-dependent observability, for example, sometimes \emph{why} an agent decides to do something is more important than  progress estimation.
Our current intention progress is a metric that measures the distance to the end of (a possible) execution trace.
However, in practice actions do not take the same length of time and further domain knowledge annotation for agent programs may be required for more realistic progress estimation. Transparency encompasses more than observability and attention directing and determining: what to log, how to log, and how to use these logs, \eg future forensic analysis or run-time decision making, will be crucial to tackle emerging HAT scenarios.
Finally,  a long-term goal  is analysis  of the accuracy of the intention progression predictions in different scenarios.
	\paragraph{\textbf{Acknowledgements}}
	This work is supported by the Engineering and Physical Sciences Research Council, under PETRAS SRF grant MAGIC (EP/S035362/1) and  S4: Science of Sensor Systems Software (EP/N007565/1).
\nocite{*}
\bibliographystyle{eptcs}
\bibliography{generic}

\end{document}